\documentclass[aps,pre,preprint,groupedaddress]{revtex4-2}

\usepackage{amsmath, amssymb, amsfonts}
\usepackage{tikz}
\begin{document}

\title{Orienting‑Field Effects on Instability and Mode Selection in Active Nematics}

\author{I.K. Joseph}
\author{A.J.H. Houston}
\author{K.N. Kowal}
\author{N.J. Mottram}
\email[Author to whom any correspondence should be addressed ]{nigel.mottram@glasgow.ac.uk}

\affiliation{School of Mathematics and Statistics, University of Glasgow, Glasgow G12 8QQ, U.K.}


\date{\today}

\begin{abstract}
We examine the instabilities of a confined active nematic subjected to an orienting field using a low Reynolds number Ericksen–Leslie framework with active stresses and field-induced torques. Linear analysis reveals two distinct modes, with odd and even director symmetry, the instabilities of which depend on the interplay between activity and field strength. We derive exact and approximate analytic forms of the stability boundaries and show that an orienting field that aligns the director perpendicular to the substrate anchoring direction cooperatively lowers activity thresholds and enables a field-driven even symmetry mode instability, while an orienting field that aligns the director parallel to the substrate anchoring tends to stabilise the system. Numerical solutions of the full nonlinear equations show that the linear stability analysis correctly identifies the symmetries of long-time states. These results demonstrate how orienting fields can promote an instability below the classical critical activity and can be used to both tune the instability onset and control the mode selection in confined active nematics.
\end{abstract}

\maketitle

\section{Introduction}\label{sec1}
Two key hallmarks of living systems are the ability to move and the ability to respond to external cues. These abilities are central to the quorum sensing of bacteria \cite{miller2001quorum}, chemotaxis \cite{parent1999cell,levine2006directional,levine2013physics,wadhams2004making} and the response of cells to mechanical stimuli \cite{persat2015mechanical}. All of these systems serve as examples of active nematics \cite{ramaswamy2010mechanics,marchetti2013hydrodynamics,doostmohammadi2018active}, non-equilibrium biological or synthetic systems composed of motile agents exhibiting orientational order in which motility is modelled by the incorporation of an active stress into the hydrodynamics of passive nematics \cite{de1993physics}. There has been great success in using the framework of active nematics to model biological processes such as apoptosis in epithelial cells \cite{saw2017topological}, the formation of colonies and biofilms by bacteria \cite{doostmohammadi2016defect,dell2018growing,basaran2022large,yaman2019emergence} and morphogenesis \cite{maroudas2021topological,guillamat2022integer}. However, this work has largely neglected the impact of external fields on the active nematic, and while this has been incorporated, for example, into models of morphogenesis \cite{wang2023patterning} and as a proxy for general anisotropic biological environments \cite{parmar2025proliferatingnematiccollectivelysenses}, there remain many gaps in the theoretical understanding of the impact of external fields.

True insight into the behaviour of biological systems, as well as many of the desired applications of active matter, require an understanding of the consequences of boundaries and confinement \cite{bechinger2016active}. This is true for active microfluidics \cite{hardouin2019reconfigurable,hardouin2020active} and the extraction of work through micro-machines \cite{ray2023rectified,houston2023colloids}, but confinement in a complex environment is also a ubiquitous biological reality and is central to collective cell dynamics \cite{saw2018biological}, biofilm formation \cite{you2021confinement} and cytoplasmic streaming \cite{woodhouse2013cytoplasmic}.
A seminal result is that of the spontaneous-flow transition \cite{voituriez2005spontaneous}, which showed that above a threshold value of activity the quiescent state becomes unstable to a distorted, flowing one, owing to the fundamental hydrodynamic instability of active nematics \cite{ramaswamy2010mechanics,simha2002hydrodynamic}. This result has since been confirmed numerically \cite{marenduzzo2007steady} and experimentally \cite{duclos2018spontaneous} and forms an essential piece of the picture of how, with increasing activity, active nematics transition from a quiescent, elasticity-dominated state, through the spontaneous-flow transition, to dancing-defect states \cite{shendruk2017dancing} and ultimately to active turbulence \cite{wensink2012meso,alert2022active}. The study of confined active nematics has since become a burgeoning field \cite{thampi2022channel}, including investigations of different geometries \cite{ravnik2013confined,opathalage2019self,alam2024active,vaidya2024active,aramini2025spontaneous}, chiral three-dimensional states \cite{keogh2022helical,pratley2024three} and the role of variations in activity \cite{houston2024spontaneous,houston2025heterogeneity}.

In this work we combine these features, studying the behaviour of an active nematic subject to both confinement and an orienting field. As an ancillary benefit, by incorporating an orienting field into a model of active nematics, our results highlight the distinctions between the activity-driven spontaneous-flow transition and the field-driven transition of passive nematics, known as the Fr\'eedericksz transition \cite{de1993physics,frisken1989freedericksz,de1971short}. Although the former has been referred to as an `active Fr\'eedericksz’ transition \cite{duclos2018spontaneous,alam2024active} it has pronounced differences in its bifurcation structure. We are able to capture the key features of the system with an analysis based on a linear stability approach - finding that the strength of the external field can be used to select the symmetry of the unstable mode and that activity and field can work cooperatively, resulting in an instability when both activity and field strength are below their classical threshold values. All of which is confirmed from the numerical solution of the full nonlinear model. Our results provide both a route to controlling active nematics \textit{in vitro} and a means by which biological systems might self-organise their collective dynamics.

\section{Model formulation}
\label{theory}
Following previous investigations \cite{walton2025orienting, walton2020mathematical}, we model an active nematic using an adapted form of the low Reynolds number Ericksen–Leslie equations for nemato-hydrodynamics \cite{stewart2019dynamic}, describing the dynamics of the director field ${\bf n}$, the average orientation of the active agents, and the velocity ${\bf v}$, by including an active component in the fluid stress tensor \cite{voituriez2005spontaneous, edwards2009spontaneous} and an orienting field torque in the balance of angular momentum \cite{walton2025orienting, joseph2025mathematical}. The orienting field may be in the form of an externally applied electric or magnetic field, which would be appropriate if the active agents exhibit dielectric or diamagnetic anisotropy, respectively, or the result of incident polarised light creating a torque due to active agent birefringence, or an induced orientation due to an anisotropic substrate that imprints a preferred director on the system. In this latter case, the model used here would assume that this substrate effect may be averaged across the dimension perpendicular to the substrate such that the orienting torque becomes a bulk term. 

We consider the simplest model in which these effects play a role, and in doing so we replicate previous models \cite{voituriez2005spontaneous,edwards2009spontaneous}, assuming a layer of active nematic  confined between two parallel plates located at $z = -d/2$ and $z = d/2$ which impose infinite planar anchoring of the director in the $x$-direction and no-slip boundary conditions for the velocity.  In stating the model below, we initially consider an orienting field that acts  to align the director towards the substrate normal, the $z$-direction, and later consider the case in which the orienting field acts to align the director perpendicular to the substrate normal, in the $x$-direction. To consider a minimal model for the instabilities, we will also assume invariance in the two in-plane directions of the layer and assume that the director remains in the plane containing the planar boundary conditions and the orienting field, the $xz$-plane. We therefore seek director solutions of the form $\mathbf{n} = (\cos\theta(z,t), 0, \sin\theta(z,t))$, so that $\theta(z, t)$ is the director angle measured with respect to the $x$-direction, and flow velocity solutions of the form $\mathbf{v} = (v(z,t),0,0)$.  

The governing equations for the flow speed and director angle are derived from the balances of linear and angular momentum, as have been described in numerous previous investigations, e.g.~\cite{walton2025orienting, walton2020mathematical}, 
\begin{align}
0=&\left(  g(\theta)v_z  + m(\theta)\theta_t - \zeta \sin \theta \cos \theta  \right)_z,\label{eqv}\\
\gamma_1 \theta_t=&  \left(K_1 \cos^2 \theta + K_3 \sin^2 \theta   \right)\theta_{zz} + (K_3 -K_1)\sin \theta \cos \theta \ \theta^2_z  - m(\theta)v_z + \chi F^2 \sin\theta \cos\theta,\label{eqth}
\end{align}
where subscripts $_t$ and $_z$ denote differentiation, $\zeta$ is the activity parameter, $\gamma_1$ is the rotational viscosity, $K_1$ and $K_3$ are the splay and bend Frank elastic constants, $\chi>0$ is a susceptibility to the effects of the orienting field and $F$ is the magnitude of the orienting field. The director dependent functions $g(\theta)$ and $m(\theta)$ are, respectively, the effective shear viscosity of the liquid crystal and the viscosity relating to the coupling between shear and director rotation, and are expressed in terms of Miesowicz viscosities  $\eta_i$ and $\gamma_i$ \cite{stewart2019dynamic} as
\begin{align}
g(\theta) & = \eta_1\cos^2\theta+\eta_2\sin^2\theta  +\eta_{12}\cos^2\theta \sin^2\theta, \label{eqg}\\
m(\theta) & = \dfrac{1}{2}\left( \gamma_1 +\gamma_2 \cos 2\theta\right).  \label{eqm}
\end{align}
The conditions of infinite planar anchoring and no-slip on the boundaries correspond to $\theta(\pm d/2,t) =0$ and $v(\pm d/2,t) =0$, respectively.
Note that a model in which the substrates impose infinite homeotropic anchoring, where the director is fixed to be parallel to the substrate normal, and the director in the bulk remains in a single plane, is in fact equivalent to the present model under the transformation $\theta\rightarrow\pi/2-\theta,\,\zeta\rightarrow -\zeta$.

A reduction of parameters may be undertaken through a non-dimensionalisation, in which we rescale time using the elastic timescale, the $z$-coordinate using the width of the region, the velocity using the typical magnitude of flow induced by rotation of the director, the orienting field using the critical field for director reorientation in the absence of activity (the critical field strength for the classical Fr\'eedericksz effect \cite{stewart2019dynamic}) and the activity parameter using the critical value for the transition to a flowing state in the absence of an orienting field. These scalings lead to non-dimensionalised time, $\hat{t}$, distance, $\hat{z}$, flow speed, $\hat{v}$, orienting field strength, $\hat{F}$ and activity, $\hat{\zeta}$, such that 
\begin{equation}
t=\dfrac{\gamma_1 d^2}{K_1}\hat{t},\quad z=d\,\hat{z},\quad v=\dfrac{2K_1}{(\gamma_1+\gamma_2) d} \hat{v},\quad F=F_c \hat{F},\quad \zeta=\zeta_c \hat{\zeta},
\end{equation}
where $F_c=\dfrac{\pi}{d}\sqrt{\dfrac{K_1}{\chi}}$ and $\zeta_c=-\dfrac{8\pi^2\eta_1K_1}{(\gamma_1+\gamma_2) d^2}$.
The resulting non-dimensionalised equations are,
\begin{align}
0=&\left(  \hat{g}(\theta)\hat{v}_{\hat{z}} + \hat{\alpha}\hat{m}(\theta)\theta_{\hat{t}} + 4\pi^2\zeta \sin \theta \cos \theta  \right)_{\hat{z}},\label{eqv2}\\
\theta_{\hat{t}}=&  \left( \cos^2 \theta + \hat{\kappa}\sin^2 \theta   \right)\theta_{\hat{z}\hat{z}} + (\hat{\kappa} -1)\sin \theta \cos \theta \ \theta^2_{\hat{z}} - \hat{m}(\theta)\hat{v}_{\hat{z}} + \pi^2\hat{F}^2 \sin\theta \cos\theta,\label{eqth2}\\
0=&  \theta(\pm 1/2,\hat{t}),\label{ndbc1}\\
0=&  \hat{v}(\pm 1/2,\hat{t})\label{ndbc2},
\end{align}
with 
\begin{align}
\hat{g}(\theta) & = \cos^2\theta+\hat{\eta}_2\sin^2\theta  +\hat{\eta}_{12}\sin^2\theta \cos^2\theta, \label{eqgnondim}\\
\hat{m}(\theta) & =1-2 \hat{\gamma}_2\sin ^2\theta,
\end{align}
where $\hat{\kappa}=K_3/K_1$ is the ratio of splay and bend elastic constants and $\hat{\eta}_2=\eta_2/\eta_1$, $\hat{\eta}_{12}=\eta_{12}/\eta_1$, $\hat{\gamma}_2=\gamma_2/(\gamma_1+\gamma_2)$ and $\hat{\alpha}=(\gamma_1+\gamma_2)^2/(4\gamma_1\eta_1)$  are ratios of viscosities. There are, therefore, seven non-dimensional parameters within this system of equations: $\hat{\zeta}$, $\hat{F}$, $\hat{\kappa}$, $\hat{\alpha}$, $\hat{\eta}_2$, $\hat{\eta}_{12}$, $\hat{\gamma}_{2}$, the first two controlling the activity and orienting field strength, which we will treat as control parameters. Since we will only use non-dimensionalised forms from now on, we will drop the $\hat{\ }$ to simplify notation.

\section{Linear perturbation about the trivial state}
\label{linear}  
Linearising about the trivial state $\theta = 0,\,v=0$, equations \eqref{eqv2} and \eqref{eqth2} become
\begin{align}
0=&v_{zz} + \alpha\theta_{t z} + 4\pi^2\zeta \theta_z, \label{eqvlinnondim}\\
\theta_t=& \theta_{zz} - v_z +\pi^2F^2 \theta,  \label{eqthlinnondim}
\end{align}
so that, apart from the control parameters associated with the orienting field and the activity parameter, there is a single material parameter, $\alpha=(\gamma_1+\gamma_2)^2/(4\gamma_1\eta_1)$, representing a ratio of viscosities. We note that for typical molecular liquid crystals we have $\alpha=0.00075$ (for MBBA), $\alpha=0.00016$ (for PAA) and $\alpha=0.00795$ (for 5CB) \cite{stewart2019dynamic}  but, to the best of our knowledge, the value for active nematics has not been measured. In later plots we set $\alpha=0.1$ simply to demonstrate the effect of this term. 
Solving eqs \eqref{eqvlinnondim} and \eqref{eqthlinnondim} together with the non-dimensionalised boundary conditions, \eqref{ndbc1} and \eqref{ndbc2}, leads to two sets of instability modes, often called the S modes and the D modes due to the shape of the director distortion of the primary mode in each case, e.g.~in \cite{pratley2024three}. The S modes, in which the director is antisymmetric about the layer mid-point, are given by
\begin{align}
&{\theta_{\rm S}}(z, t) = \bar{\theta} \left[ \sin\left(2q_{\rm S} z\right) \right] e^{\sigma_{\rm S} t}, \label{eqthsol} \\
&{v_{\rm S}}(z,t) = \bar{v} \left[ \cos\left(2q_{\rm S} z\right)- \cos q_{\rm S}\right] e^{\sigma_{\rm S} t}, \label{equsol}
\end{align}
and the D modes, in which the director is symmetric about the layer mid-point, are given by
\begin{align}
&{\theta_{\rm D}}(z, t) = \bar{\theta} \left[\cos\left(2q_{\rm D} z\right) - \cos q_{\rm D} \right] e^{\sigma_{\rm D} t}, \label{eqthsol2} \\
&{v_{\rm D}}(z, t) = \bar{v} \left[2z\sin q_{\rm D} -\sin\left(2q_{\rm D} z\right) \right] e^{\sigma_{\rm D} t}. \label{equsol2}
\end{align}
In these mode equations, $\bar{\theta}$ is a constant, determined by an initial condition, and $\bar{v} = \bar{\theta} (  4\pi^2 \zeta+\alpha\sigma_\bullet )/(2 q_\bullet)$, and where the time constants $\sigma_\bullet$ and wavenumbers $q_\bullet$ for each mode are determined, using eqs (\ref{eqvlinnondim}) and (\ref{eqthlinnondim}) together with the boundary conditions, by the equations
\begin{align}
0&=\sin q_{\rm S},\label{eqsigqS1}\\
0&=\sigma_{\rm S}\left(\alpha-1\right)-4q_{\rm S}^2+\pi^2\left(F^2+4\zeta\right),\label{eqsigqS2}
\end{align}
and
\begin{align}
0&=\sigma_{\rm D}\left(\alpha\tan q_{\rm D}-q_{\rm D}\right)+\pi^2\left(F^2q_{\rm D}+4\zeta\tan q_{\rm D}\right),\label{eqsigqD1}\\
0&=\sigma_{\rm D}(\alpha-1)-4q_{\rm D}^2+\pi^2\left(F^2+4\zeta\right).\label{eqsigqD2}
\end{align}
For the primary S mode, for which $q_{\rm S}=\pi$, the critical stability condition $\sigma_{\rm S}=0$, corresponding to the onset of instability, yields the following relationship between activity and orienting field 
\begin{align}
\zeta&=1-\dfrac{1}{4}F^2. \label{Smode}
\end{align}
For the D mode, instability occurs at the critical stability condition $\sigma_{\rm D}=0$ when the wavenumber, activity and orienting field are related by 
\begin{equation}
\zeta=-\dfrac{q^3}{\pi^2(\tan q-q)},\quad F^2=\dfrac{4q^2\tan q}{\pi^2(\tan q-q)}.
\label{Dcurve}\end{equation}
From \eqref{Dcurve} we find that the D mode critical curve in $(F,\,\zeta)$ space passes through the zero-field critical point $(F,\,\zeta)=(0,\,1)$ when $q=\pi$, through the zero-activity critical point (the classical Fr\'eedericksz transition point) $(F,\,\zeta)=(1,\,0)$ when $q=\pi/2$ and as $F$ grows larger, purely imaginary values of the wavenumber arise (from $q=0$ when $F=2\sqrt{3}/\pi$), with $|q|\rightarrow\infty$ as $F\rightarrow\infty$. 

We may obtain approximate analytical expressions for the critical curve in $(F,\,\zeta)$ space close to $q=\pi$, $q=0$ and as $|q|\rightarrow\infty$, namely
\begin{align}
\zeta_{c} &= 1 -\dfrac{3}{4}F^2 +{\rm O}(F^4),&& \text{as $F\rightarrow 0$}, \label{Dmodezerofield}\\ 
\zeta_{c} &= \dfrac{\pi^2}{4 (\pi^2-8)}(1-F^2) +{\rm O}((1-F^2)^2), && \text{as $F\rightarrow 1$}, \label{Dmodezeroactivity} \\ 
\zeta_{c} &= \dfrac{1}{\pi^2} +\dfrac{1}{4}F^2-\dfrac{\pi^2}{16}F^4  +{\rm O}\left(\dfrac{1}{F^2}\right),&& \text{as  $F\rightarrow \infty$. }\label{Dmodeinffield}
\end{align}
The last approximation is found by setting $q=iQ$, allowing $\tanh(Q)\rightarrow 1$ and asymptotically expanding for $Q\rightarrow \infty$.

In the region for which there are both S and D mode instabilities, where $\sigma_{\rm S}>0$ and  $\sigma_{\rm D}>0$, we can also find the curve in the $(F,\,\zeta)$ plane at which the S mode becomes fastest growing by considering the critical case $\sigma_{\rm S}=\sigma_{\rm D}$ in eqs \ref{eqsigqS1}-\ref{eqsigqD2}, which give  $q_{\rm S}=q_{\rm D}=\pi$, and then from \eqref{eqsigqD1} 
\begin{align}
\zeta&=1-\dfrac{\alpha}{4}F^2.\label{Smodedom}
\end{align}

\begin{figure}[ht]
 \centering
    \begin{tikzpicture}[scale=1.6]
        \node[anchor=south west,inner sep=0] at (0,0)
{\includegraphics[width=0.9\textwidth]{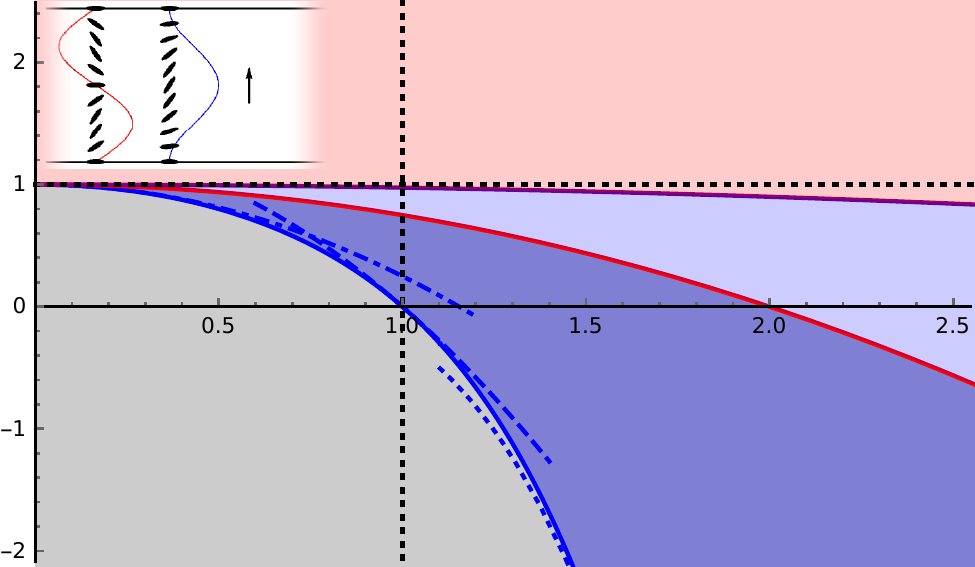}};
 \node [scale=0.75] at (0.9,5.5) {S mode};
 \node [scale=0.75] at (1.7,5.5) {D mode};
 \node [scale=1,rotate=90] at (-0.2,2.75) {activity parameter, $\zeta$};
 \node [scale=1] at (8.1,2.0) {orienting field, $F$};
 \node[scale=1,draw,fill=white,align=center] at (2,1.2) {stable trivial state};
 \node[scale=1,draw,fill=white,align=center] at (6.9,1.2) {only D mode instability};
 \node[scale=1,draw,fill=white,align=center] at (7.775,3.1) {D and S mode instabilities\\[-11pt] D mode dominates};
 \node[scale=1,draw,fill=white,align=center] at (5.8,4.6) {D and S mode instabilities\\[-11pt] S mode dominates};

\node[scale=0.75,align=left] at (2.45,5) {field in\\[-10pt] z-direction};
    \end{tikzpicture}    
\caption{Analytic critical curves for the instabilities of the primary S and D modes in a system with an orienting field that aligns the director normal to the initial director state.  The inset in the top left corner is a sketch of the director distortion and director angle profile for the S mode (red) and D mode (blue), the colours corresponding to those indicating the fastest growing mode of instability in the main plot. Blue lines indicate the D mode instability curve: the numerical solution of the parametric curve, eqs~(\ref{Dcurve}) (solid blue curve); the approximation near to the zero-field case, eq.~(\ref{Dmodezerofield}) (dashed dotted blue curve); the approximation near to the zero-activity case,  eq.~(\ref{Dmodezeroactivity}) (dashed blue curve); and the approximation as the field tends to infinity, eq.~(\ref{Dmodeinffield}) (dotted blue curve). The solid red curve is the exact analytic result for the critical  instability curve of the S mode, eq.~(\ref{Smode}). The solid purple curve indicates the boundary between regions where the D or S mode is fastest growing, eq.~(\ref{Smodedom}), for which we have used the value $\alpha=0.1$.}
\label{Fig1:analtical}
\end{figure}
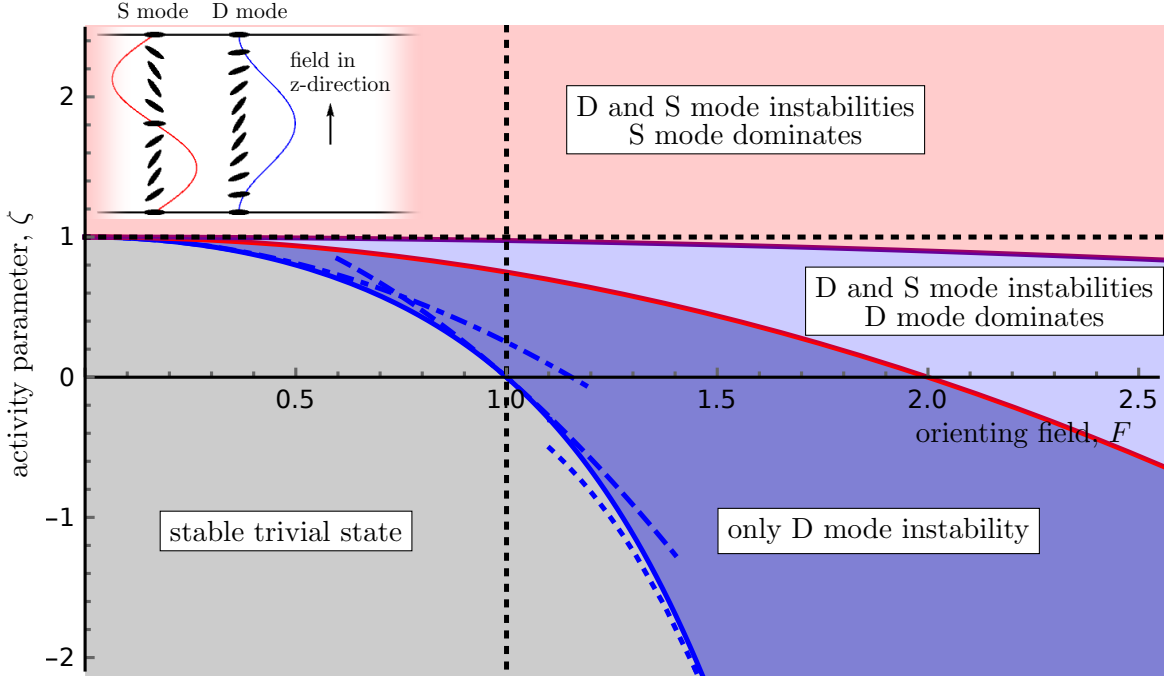

The resulting stability diagram, for the $(F,\,\zeta)$-plane, is shown in Fig.~\ref{Fig1:analtical}, constructed using the critical stability curve for the S mode, eq.~(\ref{Smode}) (solid red line), the parametric critical stability curve for the D mode in eqs~(\ref{Dcurve}) (solid blue line), the approximate forms of this parametric curve, eq.~(\ref{Dmodezerofield}) (dashed-dotted blue line), eq.~(\ref{Dmodezeroactivity})  (dashed blue line), eq.~(\ref{Dmodeinffield})  (dotted blue line) and eq.~(\ref{Smodedom}) (purple line). As expected, in this case the orienting field works against the substrate anchoring and is destabilising. We see several qualitatively distinct regions: for sufficiently small activity and field strength, to the left and below the blue lines, neither mode is unstable and the trivial planar state persists (grey region). From this stable trivial state region, increasing the field or the activity first destabilises the D mode (the dark blue region between the solid blue and solid red curves in Fig.~\ref{Fig1:analtical}), producing a Fr\'eedericksz-type instability \cite{de1993physics} in which the director distorts symmetrically about the midplane while the flow is antisymmetric. This destabilisation of the planar state by the D mode is in contrast to the zero-field case (moving from $\zeta<1$ to $\zeta>1$ along the vertical axis), the now-classical active nematic situation \cite{voituriez2005spontaneous}, where both modes destabilise at the same activity value, with the S mode being the fastest growing mode.
From the region where the D mode has been destabilised, further increasing either the field or activity destabilises the S mode, so that both symmetry modes are unstable, above the solid red curve in Fig.~\ref{Fig1:analtical}. In this doubly unstable region, mode selection is determined by the relative growth rates. In the light blue region, below the solid purple line defined by $\sigma_{\rm S}=\sigma_{\rm D}$, the D mode has the larger growth rate and therefore dominates the early-time dynamics, whereas in the red region, above the solid purple curve, the S mode is fastest growing and is expected to control the subsequent evolution.

A particularly striking feature of the stability diagram is the cooperative effect of activity and orienting field. The rectangular region in Fig.~\ref{Fig1:analtical} defined by $F<1,\,\zeta<1$ may be expected to exhibit only the trivial state, since both the activity and field are below the classical thresholds. However, regions exhibiting all the behaviours described above may be found: only D mode instability; both D and S mode instabilities, with D mode fastest growing; both D and S mode instabilities, with S mode fastest growing. In this region, activity can promote the Fr\'eedericksz transition so that director reorientation occurs at field strengths below the passive critical value, and, similarly, the presence of the field simultaneously lowers the activity threshold for instability. 

Another interesting feature of Fig.~\ref{Fig1:analtical} is that for sufficiently high orienting field, the S mode can eventually become the fastest growing mode. This may at first seem counterintuitive, because the D mode is more similar to the classic Fr\'eedericksz distortion \cite{de1993physics}, but the S mode is similar to the second-order Fr\'eedericksz distortion, and it is this mode that is being excited by the orienting field at this higher field strength. This feature gives the possibility of a field-induced selection of modes. For a particular value of the activity, $\zeta<1$, we may use eq.~(\ref{Smodedom}) to define a field strength $F=2\sqrt{(1-\zeta)\alpha}$, below which a D mode director distortion will be selected and above which a S mode director distortion will be selected. This ability for the field to select the mode will be demonstrated numerically in the next section. 

As is well known, without an orienting field, there is a double pitchfork bifurcation at the critical activity $\zeta=1$ \cite{voituriez2005spontaneous}. We see here that introducing an orienting field has removed this degeneracy and restructured the bifurcations: the field splits the onset of the S and D modes, shifting their critical activities. Previous work has also numerically demonstrated that an orienting field can stabilise or destabilise specific branches \cite{walton2025orienting}, but the current work provides analytic results for the critical curves in parameter space. The current work also shows that the field does not merely shift the zero-field threshold but qualitatively changes the symmetry and stability of the emerging states.

Using this current model we may also consider the case in which the orienting field induces a torque that tends to align the director perpendicular to the substrate normal, which in this model is parallel to the substrate anchoring direction. We would therefore expect the orienting field to stabilise the trivial planar state. For this situation, which could be due to an orienting field in the $x$-direction with $\chi>0$ or an orienting field in the $z$-direction with $\chi<0$, the only change to the governing equations is a change of sign of the $\chi F^2$ term in \eqref{eqth}. The relevant critical curves are then
\begin{align}
\zeta_c&=1+\dfrac{1}{4}F^2, \label{Smode2}
\end{align}
for the S mode instability critical curve, with the D mode critical stability curve now given parametrically by
\begin{equation}
\zeta_c=-\dfrac{q^3}{\pi^2(\tan q-q)},\quad F^2=-\dfrac{4q^2\tan q}{\pi^2(\tan q-q)}.
\label{Dcurve2}\end{equation}
Since $F^2>0$, the values of the wavenumber on the primary branch are restricted to $\pi<q<q^*$, where $q^*$ is defined by $\tan{q^*}=q^*$. The relevant approximations are then about the zero-field critical point $(F,\,\zeta)=(0,\,1)$, when $q=\pi$, and  as $F\rightarrow\infty$, when  $q\rightarrow q^*$. The first approximation is straightforward, leading to 
\begin{align}
\zeta_{c} &= 1 +\dfrac{3}{4}F^2 +{\rm O}(F^4),\quad \text{as $F\rightarrow 0$}, \label{Dmodezerofield2}
\end{align}
and for the second we expand eqs~(\ref{Dcurve2}) about $q^*$, i.e.~setting $q=q^*(1-\epsilon)$ in eqs~(\ref{Dcurve2}), expanding in $\epsilon\ll 1$, truncating at first order and then eliminating $\epsilon$ from the two resulting equations, to give
\begin{align}
\zeta_{c} &= \left(\dfrac{q^*}{\pi}\right)^2+\dfrac{1}{4} F^2+{\rm O}\left(\dfrac{1}{F^2}\right),\quad \text{as $F\rightarrow \infty$}. \label{Dmodeinffield2}
\end{align}

\begin{figure}[ht]
 \centering
    \begin{tikzpicture}[scale=1.6]
        \node[anchor=south west,inner sep=0] at (0,0)
{\includegraphics[width=0.9\textwidth]{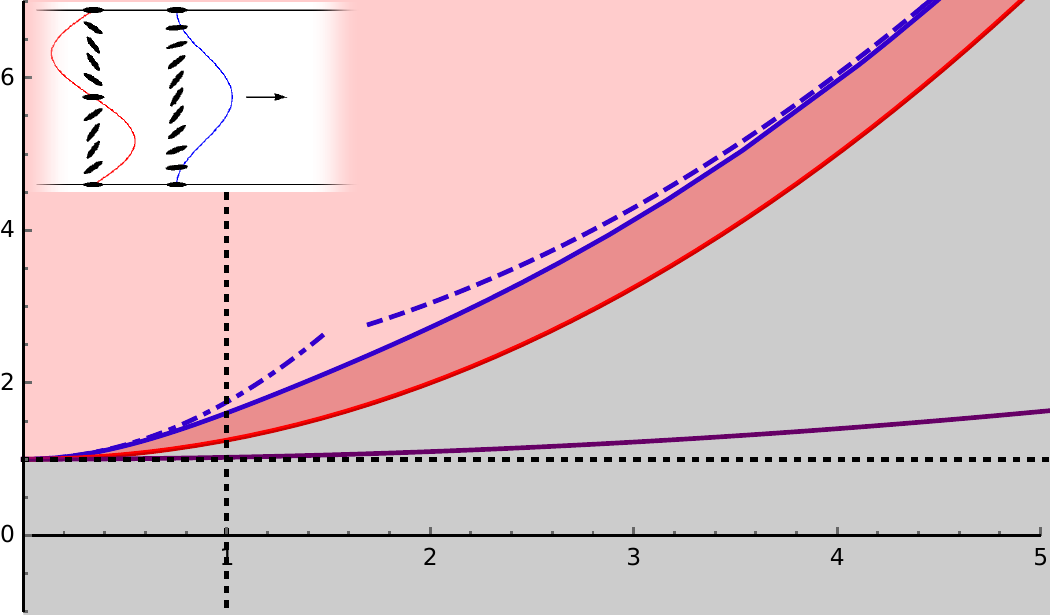}};
 \node [scale=0.75] at (0.9,5.55) {S mode};
 \node [scale=0.75] at (1.7,5.55) {D mode};
 \node [scale=1,rotate=90] at (-0.2,2.75) {activity parameter, $\zeta$};
 \node [scale=1] at (8.1,0.25) {orienting field, $F$};
 \node[scale=1,draw,fill=white,align=center] at (4.3,1.0)  {stable trivial state};
 \node[scale=1,draw,fill=white,align=center] at (7,2.7)  (box) {only S mode instability};
 \draw[->, thick] (box.north west) -- ++(-0.2,0.2) node[right] {};
 \node[scale=1,draw,fill=white,align=center] at (1.95,3.1) {D and S mode instabilities\\[-10pt] S mode dominates};

 \node[scale=0.75,align=left] at (2.6,4.8) {field in\\[-10pt] x-direction};

    \end{tikzpicture}    
\caption{Analytic critical curves for the instabilities of the primary S and D modes in a system with an orienting field that aligns the director parallel to the substrate director anchoring state. The inset in the top left corner is a sketch of the director distortion and director angle profile for the S mode (red) and D mode (blue), the colours corresponding to those indicating the fastest growing unstable mode in the main plot.  
Blue lines indicate the D mode instability curve: the numerical solution of the parametric curve, eqs~(\ref{Dcurve}) (solid blue curve); the approximation near to the zero-field case, eq.~(\ref{Dmodezerofield2}) (dashed dotted blue curve); and the approximation as the field tends to infinity, eq.~(\ref{Dmodeinffield2}) (dotted blue curve). The solid red curve is the exact analytic result for the critical  instability curve of the S mode, eq.~(\ref{Smode2}). The solid purple curve indicates where the D and S mode have the same time constant, eq.~(\ref{Smodedom2}), for which we have used the value $\alpha=0.1$.}
\label{Fig2:analticalneg}
\end{figure}
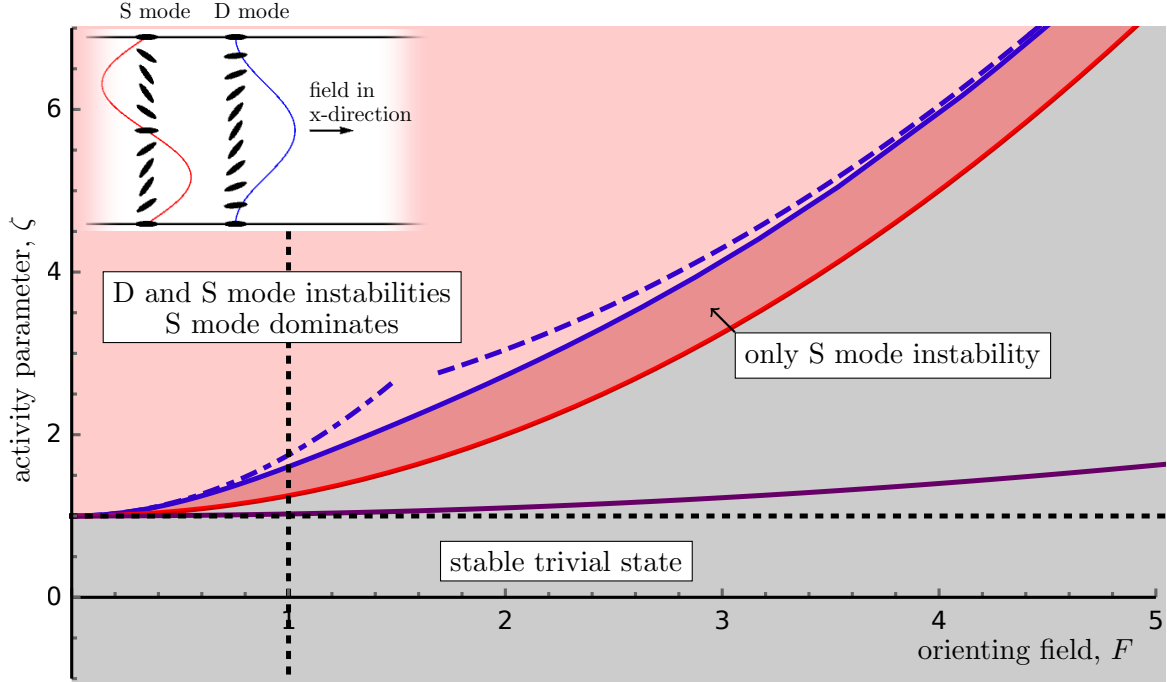

We can also find the curve in the $(F,\,\zeta)$ plane at which the D and S modes have the same time constant by considering the case $\sigma_{\rm S}=\sigma_{\rm D}$ and setting $q_{\rm S}=q_{\rm D}=\pi$. From \eqref{eqsigqD1} we then have
\begin{equation}
\zeta=1+\dfrac{\alpha}{4}F^2,\label{Smodedom2}
\end{equation}
although since this curve is entirely within the region in which $\sigma_{\rm S}<0,\,\sigma_{\rm D}<0$ and so the trivial state is stable, it does not lead to any change in the system stability. 

As would be expected, the behaviour changes qualitatively when the orienting field tends to align the director parallel  to the anchoring direction, since the field is now stabilising the undistorted planar state.  The corresponding stability diagram, shown in Fig.~\ref{Fig2:analticalneg}, shows that for any activity, a sufficiently high orienting field will stabilise the trivial planar state. Reducing the orienting field, or increasing the activity, will lead to an instability of the S mode (the dark red region above the solid red line in Fig.~\ref{Fig2:analticalneg}) and eventually the D mode (above the solid blue line in Fig.~\ref{Fig2:analticalneg}), although the S mode is always fastest growing in this region since it is above the curve defined in eq.~(\ref{Smodedom2}).

\section{Numerical solution of the full nonlinear system} 
\label{numerical}  

The analytical results presented in the previous section provide insight into the onset of instability and the symmetry of emerging modes, but do not fully determine the nonlinear evolution of the system. In particular, while the reduced analysis predicts the appearance of symmetric and antisymmetric director distortions, it does not establish which states are dynamically selected at late times or how their stability depends on the control parameters. To address these questions, we investigate the full nonlinear dynamics using direct numerical simulations in COMSOL Multiphysics \cite{COMSOL:2025}. In all simulations, the same initial condition is used, a small perturbation ($||\theta||_2,\,||u||_2\approx 0.05$) in the form of a mix of S and D modes. The default COMSOL settings are used in these simulations, with adaptive time-stepping, a spatial mesh of 100 quadratic finite elements and a non-dimensional run-time of $t=4$, all of which is sufficient to ensure that further refinement and longer run-times do not change the $L^2$-norm of the solutions by more that 0.1\%.    

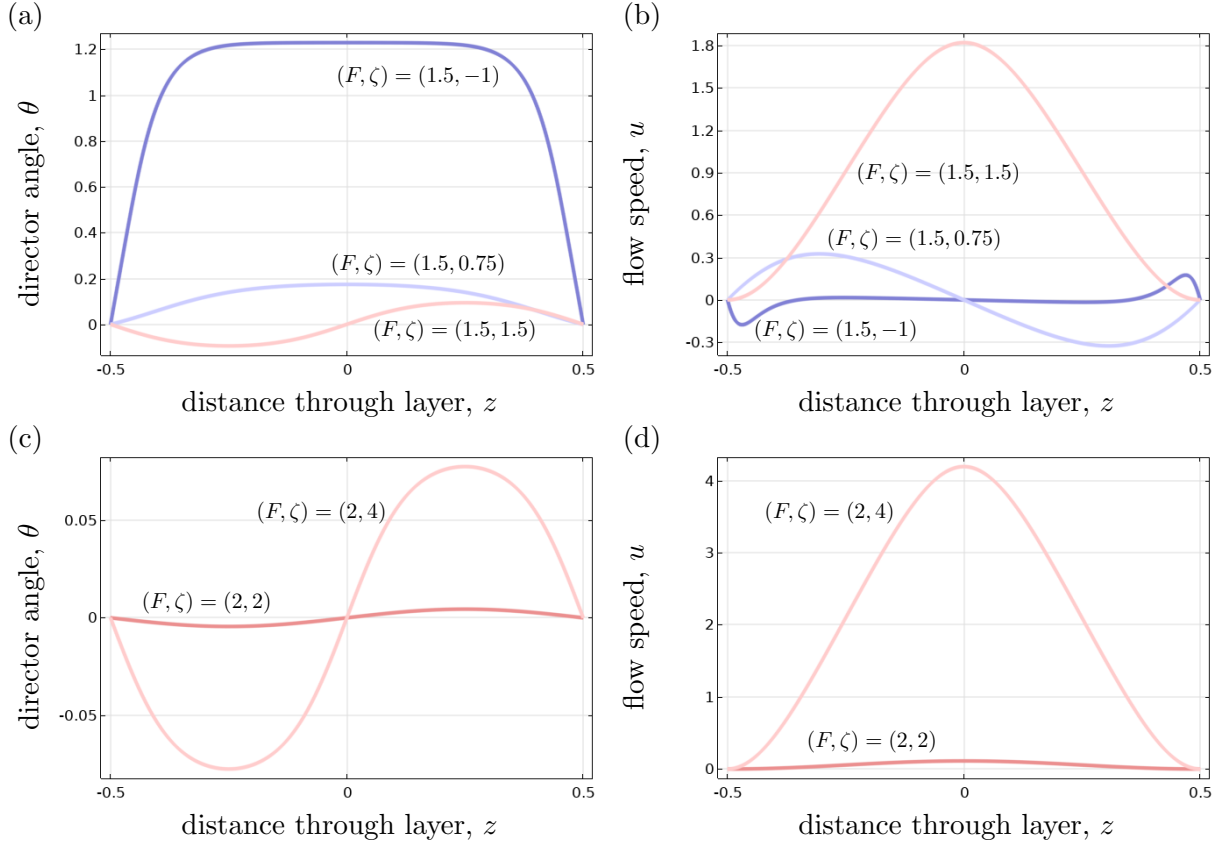
\begin{figure}[ht]
 \centering
    \begin{tikzpicture}[scale=1.6]
\node[anchor=south west,inner sep=0] at (-0.4,3.5)
{\includegraphics[width=0.45\textwidth]{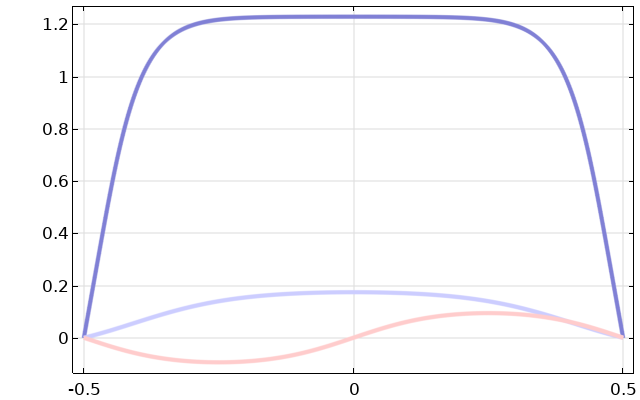}};
\node[anchor=south west,inner sep=0] at (4.7,3.5)
{\includegraphics[width=0.45\textwidth]{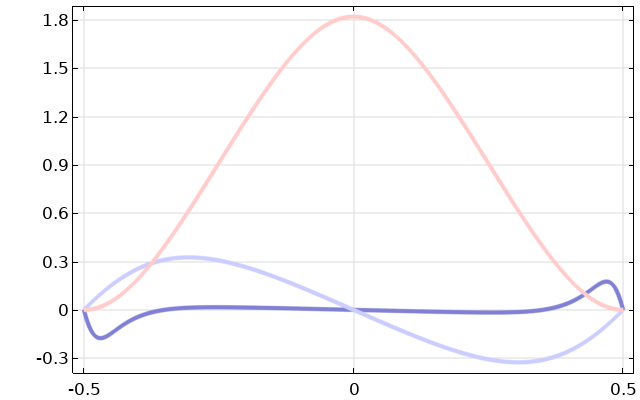}};
\node[anchor=south west,inner sep=0] at (-0.4,0)
{\includegraphics[width=0.45\textwidth]{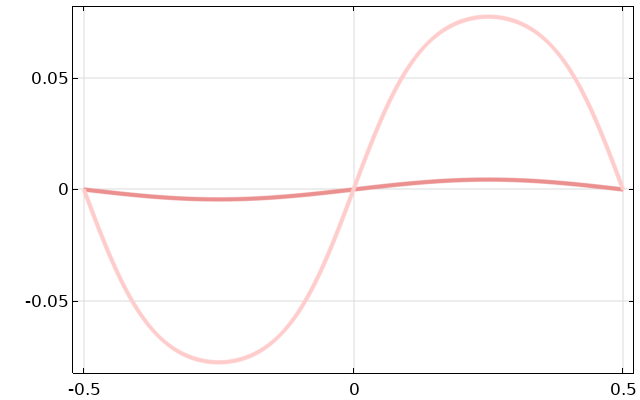}};
\node[anchor=south west,inner sep=0] at (4.7,0)
{\includegraphics[width=0.45\textwidth]{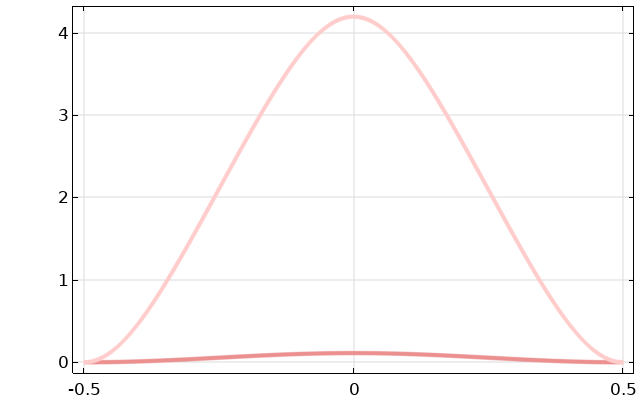}};
 \node [scale=1,rotate=90] at (-0.45,4.95) {director angle, $\theta$};
 \node [scale=1,rotate=90] at (4.55,4.95) {flow speed, $u$};
 \node [scale=1] at (2.1,3.3) {distance through layer, $z$};
 \node [scale=1] at (7.2,3.3) {distance through layer, $z$};
 \node [scale=1,rotate=90] at (-0.45,1.45) {director angle, $\theta$};
 \node [scale=1,rotate=90] at (4.55,1.45) {flow speed, $u$};
 \node [scale=1] at (2.1,-0.2) {distance through layer, $z$};
 \node [scale=1] at (7.2,-0.2) {distance through layer, $z$};
 \node[scale=1,align=center] at (-0.5,6.5)  {(a)};
 \node[scale=1,align=center] at (4.6,6.5)  {(b)};
 \node[scale=1,align=center] at (-0.5,3)  {(c)};
 \node[scale=1,align=center] at (4.6,3)  {(d)};
 \node[scale=0.75,align=left] at (2.75,6)  {$(F,\zeta)=(1.5,-1)$};
 \node[scale=0.75,align=left] at (2.75,4.45)  {$(F,\zeta)=(1.5,0.75)$};
 \node[scale=0.75,align=left] at (3.05,3.9)  {$(F,\zeta)=(1.5,1.5)$};
 \node[scale=0.75,align=left] at (6.2,3.9)  {$(F,\zeta)=(1.5,-1)$};
 \node[scale=0.75,align=left] at (6.85,4.65)  {$(F,\zeta)=(1.5,0.75)$};
 \node[scale=0.75,align=left] at (7.05,5.2)  {$(F,\zeta)=(1.5,1.5)$};
 \node[scale=0.75,align=left] at (1.95,2.4)  {$(F,\zeta)=(2,4)$};
 \node[scale=0.75,align=left] at (1,1.65)  {$(F,\zeta)=(2,2)$};
 \node[scale=0.75,align=left] at (6.15,2.4)  {$(F,\zeta)=(2,4)$};
 \node[scale=0.75,align=left] at (6.5,0.5)  {$(F,\zeta)=(2,2)$};
    \end{tikzpicture}    
\caption{Director angle and flow speed profiles for various values of the non-dimensional orienting field strength, $F$, and activity, $\zeta$, parameters. In (a) and (b) the orienting field is destabilising, with $(F,\zeta)=(1.5,-1)$ (dark blue),  $(1.5,0.75)$ (light blue) and  $(1.5,1.5)$ (light red). In (c) and (d) the orienting field is stabilising, with $(F,\zeta)=(2,2)$ (light red) and  $(2,4)$ (dark red). Colours correspond to the shading of areas in Figs \ref{Fig1:analtical} and \ref{Fig2:analticalneg}.}
\label{Fig3:numerical_states}
\end{figure}

\begin{figure}[ht]
 \centering
    \begin{tikzpicture}[scale=1.6]
\node[anchor=south west,inner sep=0] at (-0.4,0)
{\includegraphics[width=0.45\textwidth]{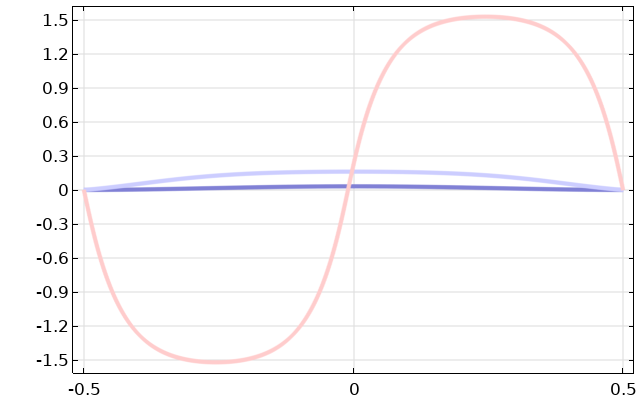}};
\node[anchor=south west,inner sep=0] at (4.7,0)
{\includegraphics[width=0.45\textwidth]{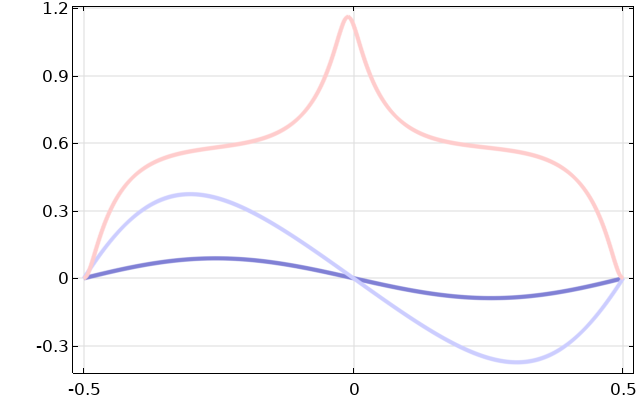}};
 \node [scale=1,rotate=90] at (-0.45,1.45) {director angle, $\theta$};
 \node [scale=1,rotate=90] at (4.55,1.45) {flow speed, $u$};
 \node [scale=1] at (2.1,-0.2) {distance through layer, $z$};
 \node [scale=1] at (7.2,-0.2) {distance through layer, $z$};
 \node[scale=1,align=center] at (-0.5,3)  {(a)};
 \node[scale=1,align=center] at (4.6,3)  {(b)};
 \node[scale=0.75,align=left] at (1.7,2.5)  {$(F,\zeta)=(10,0.9)$};
 \node[scale=0.75,align=left] at (1,1.8)  {$(F,\zeta)=(1.5,0.9)$};
 \node[scale=0.75,align=left] at (1,1.35)  {$(F,\zeta)=(0.5,0.9)$};
 \node[scale=0.75,align=left] at (6.4,2.5)  {$(F,\zeta)=(10,0.9)$};
 \node[scale=0.75,align=left] at (7.15,1.5)  {$(F,\zeta)=(1.5,0.9)$};
 \node[scale=0.75,align=left] at (6.2,0.775)  {$(F,\zeta)=(0.5,0.9)$};
    \end{tikzpicture}    
\caption{Demonstration of cooperative instability and mode selection by varying field strength with a fixed activity, for the case in which the orienting field is destabilising.
Director angle and flow speed profiles are shown in (a) and (b), respectively, for a fixed activity, $\zeta=0.9$, which is below the critical value in the absence of an orienting field. Profiles are shown for varying values of the orienting field, $F$: $F=0.5$ (dark blue), below the classical Fr\'eedericksz value, where field and activity are working cooperatively; above the Fr\'eedericksz value where the orienting field is chosen to select the D mode,  $F=1.5$ (light blue); above the Fr\'eedericksz value  where the orienting field is chosen to select the S mode, $F=10$ (light red). Colours correspond to the shading of areas in Fig.~\ref{Fig1:analtical}.}
\label{Fig4:fieldselection}
\end{figure}

We therefore now examine representative steady states of the nonlinear system at selected points in parameter space. Cases are chosen from the distinct regions of the analytical phase diagrams of Figs \ref{Fig1:analtical} and \ref{Fig2:analticalneg}, and are shown in Fig.~\ref{Fig3:numerical_states}.  In Fig.~\ref{Fig3:numerical_states}(a) and (b) we plot the director angle and flow speed profiles, respectively, for the case of a destabilising orienting field. These profiles are for: $(F,\,\zeta)=(1.5,\,-1)$, within the dark blue region of Fig.~\ref{Fig1:analtical}, where the D mode instability occurs; $(F,\,\zeta)=(1.5,\,0.75)$, within the light blue region of Fig.~\ref{Fig1:analtical}, where both D and S mode instabilities occur but we see that the D mode is selected, as predicted from the linear growth rates; and $(F,\,\zeta)=(1.5,\,1.5)$, within the light red region of Fig.~\ref{Fig1:analtical}, where both D and S mode instabilities occur but we see that the S mode is selected, again as predicted from the linear growth rates. In Fig.~\ref{Fig3:numerical_states}(c) and (d) we plot the director angle and flow speed profiles, respectively, for the case of a stabilising orienting field. These profiles are for: $(F,\,\zeta)=(2,\,2)$, within the dark red region of Fig.~\ref{Fig2:analticalneg}, where the S mode instability occurs; and $(F,\,\zeta)=(2,\,4)$, within the light red region of Fig.~\ref{Fig2:analticalneg}, where both D and S mode instabilities occur but we see that the S mode is selected, as predicted from the linear growth rates.
These numerical profiles therefore demonstrate that the long-term behaviour matches the predicted linear instabilities.    

In Fig.~\ref{Fig4:fieldselection} we confirm the features predicted by the linear stability analysis described in the previous section, that an instability can occur at activity and orienting field values both below their classical critical values and that the orienting field is able to select the long-term symmetry of the system. We consider the long-time director angle and flow speed profiles at a fixed activity, $\zeta=0.9$, which is below the classical zero-field threshold, for three values of orienting field. The profiles for $(F,\zeta)=(0.5,0.9)$ (dark blue lines) demonstrate that a distorted director profile is stable even below the classical zero-field activity threshold and passive field threshold. This state demonstrates that an orienting field and activity can work cooperatively to destabilise the system. The profiles for $(F,\zeta)=(1.5,0.9)$ (light blue lines) and $(F,\zeta)=(10,0.9)$ (light red lines)  demonstrate the ability for the orienting field to select the long-term symmetry of the system, at a fixed value of activity -- selecting either the D mode, for low field strengths, or the S mode, for high field strengths. 

\section{Conclusion}
In this work we have analysed how an orienting field modifies instability onset and mode selection in a confined active nematic. Using a minimal Ericksen–Leslie framework augmented by active stresses and a field‑induced torque, we combined linear stability analysis with nonlinear numerical simulations to clarify the distinct and cooperative roles played by activity, confinement, and external alignment cues.

The linear analysis reveals the structure of the instabilities - that an orienting field generically breaks the degeneracy between the S (antisymmetric director) and D (symmetric director) instability modes that characterise the zero‑field spontaneous-flow transition. When the orienting field favours alignment normal to the confining plates, it not only shifts the classical activity threshold but introduces a field‑driven D mode instability and creates extended regions of parameter space in which both modes coexist with different growth rates. In this regime, activity and field act cooperatively: activity can promote a Fr\'eedericksz‑type reorientation below the passive critical field strength, while the field simultaneously lowers the activity required for instability. By contrast, when the field aligns the director parallel to the anchoring direction, it suppresses instabilities and stabilises the trivial state over a wide range of activity, ensuring that any instability that does occur is dominated by the S mode.

Direct numerical simulations of the fully nonlinear equations confirm that the linear instability boundaries reliably predict the long‑time behaviour of the system. The simulations demonstrate that the orienting field provides a robust and continuous control parameter for selecting the symmetry of the emerging director distortion and flow profile at fixed activity, enabling transitions between symmetric and antisymmetric steady states without changing material properties or confinement. Importantly, these symmetry‑selecting effects persist even when both the activity and field lie below their classical passive thresholds.

Taken together, our results show that orienting fields do far more than simply stabilise or destabilise an active nematic: they qualitatively restructure the bifurcation landscape, remove the mode degeneracy, and enable controlled selection of flow states in confinement. This provides a clear physical framework for interpreting possible future experiments on field‑responsive and substrate‑guided active nematics, and suggests practical strategies for designing reconfigurable active‑matter devices. Extensions of this work to finite anchoring, patterned fields, and higher‑dimensional director structures \cite{alam2024active,pratley2024three} offer promising directions for future study.

\acknowledgments{
\noindent For the purpose of open access, the authors have applied a Creative
\noindent Commons Attribution (CC-BY) licence to any Author Accepted Manuscript
version arising from this submission.}

\noindent I.J. and N.M. would like to acknowledge the financial support of the University of Glasgow, and N.M. would like to acknowledge the financial support of the Medical Research Council [grant number G0902331].

\noindent {N.M. and K.K. conceived the research and methodology, I.J., A.H. and N.M. carried out the calculations and analysis, I.J. and N.M. carried out the numerical work, I.J. and N.M. produced an initial version of the manuscript and all authors contributed to the editing and final version.}

\noindent {No new data were created or analysed in this study. The COMSOL Multiphysics code used to generate the numerical profiles in Figs \ref{Fig3:numerical_states} and \ref{Fig4:fieldselection} is available on request.}

\bibliography{JosephHoustonKowalMottram_2026_biblio}

@book{stewart2019dynamic,
  title={Static and Dynamic Continuum Theory of Liquid Crystals: A Mathematical Introduction},
  author={Stewart, I.W.},
  year={2004},
  publisher={Taylor \& Francis Group}
}

@book{de1993physics,
  title={The Physics of Liquid Crystals},
  author={de Gennes, P.G. and Prost, J.},
  year={1993},
  publisher={Oxford University Press}
}

@phdthesis{hardouin2020active,
  title={Active Liquid Crystals in Confinement},
  author={Hardo{\"u}in, J.},
  year={2020},
  school={Universitat de Barcelona}
}

@phdthesis{walton2020mathematical,
  title={Mathematical Modelling of Active Nematic Liquid Crystals in Confined Regions},
  author={Walton, J.},
  year={2020},
  school={University of Strathclyde}
}

@article{shendruk2017dancing,
  title={Dancing Disclinations in Confined Active Nematics},
  author={Shendruk, T.N. and Doostmohammadi, A. and Thijssen, K. and Yeomans, J.M.},
  journal={Soft Matter},
  volume={13},
  number={21},
  pages={3853--3862},
  year={2017},
  publisher={Royal Society of Chemistry}
}

@article{edwards2009spontaneous,
  title={Spontaneous flow states in active nematics: a unified picture},
  author={Edwards, S.A. and Yeomans, J.M.},
  journal={Europhysics Letters},
  volume={85},
  number={1},
  pages={18008},
  year={2009}
}

@article{wensink2012meso,
  title={Meso-scale turbulence in living fluids},
  author={Wensink, H.H. and Dunkel, J. and Heidenreich, S. and Drescher, K. and Goldstein, R.E. and L{\"o}wen, H. and Yeomans, J.M.},
  journal={Proceedings of the National Academy of Sciences},
  volume={109},
  number={36},
  pages={14308--14313},
  year={2012}
}

@article{ravnik2013confined,
  title={Confined active nematic flow in cylindrical capillaries},
  author={Ravnik, M. and Yeomans, J.M.},
  journal={Physical Review Letters},
  volume={110},
  number={2},
  pages={026001},
  year={2013}
}

@article{voituriez2005spontaneous,
  title={Spontaneous flow transition in active polar gels},
  author={Voituriez, R. and Joanny, J.F. and Prost, J.},
  journal={Europhysics Letters},
  volume={70},
  number={3},
  pages={404},
  year={2005}
}

@article{marchetti2013hydrodynamics,
  title={Hydrodynamics of soft active matter},
  author={Marchetti, M.C. and Joanny, J. and Ramaswamy, S. and Liverpool, T.B. and Prost, J. and Rao, M. and Simha, R.A.},
  journal={Reviews of Modern Physics},
  volume={85},
  number={3},
  pages={1143--1189},
  year={2013}
}

@article{thampi2022channel,
  title={Channel confined active nematics},
  author={Thampi, S.P.},
  journal={Current Opinion in Colloid \& Interface Science},
  volume={61},
  pages={101613},
  year={2022},
  publisher={Elsevier}
}

@article{duclos2018spontaneous,
  title={Spontaneous shear flow in confined cellular nematics},
  author={Duclos, G. and Blanch-Mercader, C. and Yashunsky, V. and Salbreux, G. and Joanny, J.-F. and Prost, J. and Silberzan, P.},
  journal={Nature Physics},
  volume={14},
  number={7},
  pages={728--732},
  year={2018},
  publisher={Nature Publishing Group}
}

@article{marenduzzo2007steady,
  title={Steady-state hydrodynamic instabilities of active liquid crystals: Hybrid lattice Boltzmann simulations},
  author={Marenduzzo, D. and Orlandini, E. and Cates, M.E. and Yeomans, J.M.},
  journal={Physical Review E},
  volume={76},
  number={3},
  pages={031921},
  year={2007},
  publisher={APS}
}

@article{hardouin2019reconfigurable,
  title     = {Reconfigurable flows and defect landscape of confined active nematics},
  author    = {Hardo{\"u}in, J. and Hughes, R. and Doostmohammadi, A. and Laurent, J. and Lopez-Leon, T. and Yeomans, J.M. and Ign{\'e}s-Mullol, J. and Sagu{\'e}s, F.},
  journal   = {Communications Physics},
  volume    = {2},
  number    = {1},
  pages     = {121},
  year      = {2019},
  publisher = {Nature Publishing Group UK London}
}

@article{doostmohammadi2018active,
  title={Active nematics},
  author={Doostmohammadi, A. and Ign{\'e}s-Mullol, J. and Yeomans, J.M. and Sagu{\'e}s, F.},
  journal={Nature Communications},
  volume={9},
  number={1},
  pages={3246},
  year={2018},
  publisher={Nature Publishing Group}
}

@article{simha2002hydrodynamic,
  title={Hydrodynamic fluctuations and instabilities in ordered suspensions of self-propelled particles},
  author={Simha, R.A. and Ramaswamy, S.},
  journal={Physical Review Letters},
  volume={89},
  number={5},
  pages={058101},
  year={2002},
  publisher={APS}
}

@manual{COMSOL:2025,
  title={COMSOL Multiphysics Version 6.3},
  author={{COMSOL, Inc.}},
  year={2025},
  address={Burlington, MA, USA},
  url={http://www.comsol.com}
}

@article{ramaswamy2010mechanics,
  title={The mechanics and statistics of active matter},
  author={Ramaswamy, S.},
  journal={Annual Review of Condensed Matter Physics},
  volume={1},
  number={1},
  pages={323--345},
  year={2010},
  publisher={Annual Reviews}
}

@article{you2021confinement,
  title={Confinement-induced self-organization in growing bacterial colonies},
  author={You, Z. and Pearce, D.J.G. and Giomi, L.},
  journal={Science Advances},
  volume={7},
  number={4},
  pages={eabc8685},
  year={2021},
  publisher={American Association for the Advancement of Science}
}

@article{yaman2019emergence,
  title={Emergence of active nematics in chaining bacterial biofilms},
  author={Yaman, Y.I. and Demir, E. and Vetter, R. and Kocabas, A.},
  journal={Nature Communications},
  volume={10},
  number={1},
  pages={2285},
  year={2019},
  publisher={Nature Publishing Group UK London}
}

@article{opathalage2019self,
  title={Self-organized dynamics and the transition to turbulence of confined active nematics},
  author={Opathalage, A. and Norton, M.M. and Juniper, M.P.N. and Langeslay, B. and Aghvami, S.A. and Fraden, S. and Dogic, Z.},
  journal={Proceedings of the National Academy of Sciences},
  volume={116},
  number={11},
  pages={4788--4797},
  year={2019},
  publisher={National Academy of Sciences}
}

@article{alam2024active,
  title={Active Fr{\'e}edericksz transition in active nematic droplets},
  author={Alam, S. and Najma, B. and Singh, A. and Laprade, J. and Gajeshwar, G. and Yevick, H.G. and Baskaran, A. and Foster, P.J. and Duclos, G.},
  journal={Physical Review X},
  volume={14},
  number={4},
  pages={041002},
  year={2024},
  publisher={APS}
}

@article{pratley2024three,
  title={Three-dimensional spontaneous flow transition in a homeotropic active nematic},
  author={Pratley, V.J. and Caf, E. and Ravnik, M. and Alexander, G.P.},
  journal={Communications Physics},
  volume={7},
  number={1},
  pages={127},
  year={2024},
  publisher={Nature Publishing Group UK London}
}

@article{de1971short,
  title={Short range order effects in the isotropic phase of nematics and cholesterics},
  author={de Gennes, P. G.},
  journal={Molecular Crystals and Liquid Crystals},
  volume={12},
  number={3},
  pages={193--214},
  year={1971},
  publisher={Taylor \& Francis}
}

@phdthesis{joseph2025mathematical,
  author       = {Joseph, Ijuptil Kwajighu},
  title        = {Mathematical Modelling of Active Fluids in a Channel},
  school       = {University of Glasgow},
  year         = {2025}
}

@article{parmar2025proliferatingnematiccollectivelysenses,
      title={A Proliferating Nematic That Collectively Senses an Anisotropic Substrate}, 
      author={Toshi Parmar and Fridtjof Brauns and Yimin Luo and M. Cristina Marchetti},
      year={2026},
      journal={PRX Life},
  volume={4},
  pages={013010}, 
}

@article{walton2025orienting,
  title={Orienting field effects on the flow of an active nematic liquid crystal in a channel},
  author={Walton, J. and McKay, G. and Mottram, N.J.},
  journal={European Physical Journal E},
  volume={48},
  number={10},
  pages={67},
  year={2025},
  publisher={Springer}
}

@article{frisken1989freedericksz,
  title={Freedericksz transitions in nematic liquid crystals: The effects of an in-plane electric field},
  author={Frisken, B.J. and Palffy-Muhoray, P.},
  journal={Physical Review A},
  volume={40},
  number={10},
  pages={6099},
  year={1989},
  publisher={APS}
}

@article{miller2001quorum,
  title={Quorum sensing in bacteria},
  author={Miller, Melissa B and Bassler, Bonnie L},
  journal={Annual Review of Microbiology},
  volume={55},
  number={1},
  pages={165--199},
  year={2001},
  publisher={Annual Reviews 4139 El Camino Way, PO Box 10139, Palo Alto, CA 94303-0139, USA}
}

@article{parent1999cell,
  title={A cell's sense of direction},
  author={Parent, Carole A and Devreotes, Peter N},
  journal={Science},
  volume={284},
  number={5415},
  pages={765--770},
  year={1999},
  publisher={American Association for the Advancement of Science}
}

@article{levine2006directional,
  title={Directional sensing in eukaryotic chemotaxis: a balanced inactivation model},
  author={Levine, Herbert and Kessler, David A and Rappel, Wouter-Jan},
  journal={Proceedings of the National Academy of Sciences},
  volume={103},
  number={26},
  pages={9761--9766},
  year={2006},
  publisher={National Academy of Sciences}
}

@article{levine2013physics,
  title={The physics of eukaryotic chemotaxis},
  author={Levine, Herbert and Rappel, Wouter-Jan},
  journal={Physics Today},
  volume={66},
  number={2},
  pages={24--30},
  year={2013},
  publisher={AIP Publishing}
}

@article{wadhams2004making,
  title={Making sense of it all: bacterial chemotaxis},
  author={Wadhams, George H and Armitage, Judith P},
  journal={Nature Reviews Molecular Cell Biology},
  volume={5},
  number={12},
  pages={1024--1037},
  year={2004},
  publisher={Nature Publishing Group UK London}
}

@article{persat2015mechanical,
  title={The mechanical world of bacteria},
  author={Persat, Alexandre and Nadell, Carey D and Kim, Minyoung Kevin and Ingremeau, Francois and Siryaporn, Albert and Drescher, Knut and Wingreen, Ned S and Bassler, Bonnie L and Gitai, Zemer and Stone, Howard A},
  journal={Cell},
  volume={161},
  number={5},
  pages={988--997},
  year={2015},
  publisher={Elsevier}
}

@article{saw2017topological,
  title={Topological defects in epithelia govern cell death and extrusion},
  author={Saw, Thuan Beng and Doostmohammadi, Amin and Nier, Vincent and Kocgozlu, Leyla and Thampi, Sumesh and Toyama, Yusuke and Marcq, Philippe and Lim, Chwee Teck and Yeomans, Julia M and Ladoux, Benoit},
  journal={Nature},
  volume={544},
  number={7649},
  pages={212--216},
  year={2017},
  publisher={Nature Publishing Group}
}

@article{doostmohammadi2016defect,
  title={Defect-mediated morphologies in growing cell colonies},
  author={Doostmohammadi, Amin and Thampi, Sumesh P and Yeomans, Julia M},
  journal={Physical Review Letters},
  volume={117},
  number={4},
  pages={048102},
  year={2016},
  publisher={APS}
}

@article{dell2018growing,
  title={A growing bacterial colony in two dimensions as an active nematic},
  author={Dell’Arciprete, Dario and Blow, Matthew L and Brown, Aidan T and Farrell, Fred DC and Lintuvuori, Juho S and McVey, Alexander F and Marenduzzo, Davide and Poon, Wilson CK},
  journal={Nature Communications},
  volume={9},
  number={1},
  pages={4190},
  year={2018},
  publisher={Nature Publishing Group UK London}
}

@article{basaran2022large,
  title={Large-scale orientational order in bacterial colonies during inward growth},
  author={Basaran, Mustafa and Yaman, Y Ilker and Y{\"u}ce, Tevfik Can and Vetter, Roman and Kocabas, Askin},
  journal={eLife},
  volume={11},
  pages={e72187},
  year={2022},
  publisher={eLife Sciences Publications, Ltd}
}

@article{maroudas2021topological,
  title={Topological defects in the nematic order of actin fibres as organization centres of Hydra morphogenesis},
  author={Maroudas-Sacks, Yonit and Garion, Liora and Shani-Zerbib, Lital and Livshits, Anton and Braun, Erez and Keren, Kinneret},
  journal={Nature Physics},
  volume={17},
  number={2},
  pages={251--259},
  year={2021},
  publisher={Nature Publishing Group UK London}
}

@article{guillamat2022integer,
  title={Integer topological defects organize stresses driving tissue morphogenesis},
  author={Guillamat, Pau and Blanch-Mercader, Carles and Pernollet, Guillaume and Kruse, Karsten and Roux, Aur{\'e}lien},
  journal={Nature Materials},
  volume={21},
  number={5},
  pages={588--597},
  year={2022},
  publisher={Nature Publishing Group UK London}
}

@article{wang2023patterning,
  title={Patterning of morphogenetic anisotropy fields},
  author={Wang, Zihang and Marchetti, M Cristina and Brauns, Fridtjof},
  journal={Proceedings of the National Academy of Sciences},
  volume={120},
  number={13},
  pages={e2220167120},
  year={2023},
  publisher={National Academy of Sciences}
}

@article{houston2024spontaneous,
  title={Spontaneous flows and quantum analogies in heterogeneous active nematic films},
  author={Houston, Alexander J H and Mottram, Nigel J},
  journal={Communications Physics},
  volume={7},
  number={1},
  pages={375},
  year={2024},
  publisher={Nature Publishing Group UK London}
}

@misc{houston2025heterogeneity,
      title={Heterogeneity-Induced Oscillations in Active Nematics}, 
      author={Alexander J. H. Houston and Michael Grinfeld and Geoff McKay and Nigel J. Mottram},
      journal={},
      eprint={2512.11123},
      archivePrefix={arXiv},
      primaryClass={cond-mat.soft},
      url={https://arxiv.org/abs/2512.11123},
      year={2025}
}

@article{alert2022active,
  title={Active turbulence},
  author={Alert, Ricard and Casademunt, Jaume and Joanny, Jean-Fran{\c{c}}ois},
  journal={Annual Review of Condensed Matter Physics},
  volume={13},
  number={1},
  pages={143--170},
  year={2022},
  publisher={Annual Reviews}
}

@article{keogh2022helical,
  title={Helical flow states in active nematics},
  author={Keogh, Ryan R and Chandragiri, Santhan and Loewe, Benjamin and Ala-Nissila, Tapio and Thampi, Sumesh P and Shendruk, Tyler N},
  journal={Physical Review E},
  volume={106},
  number={1},
  pages={L012602},
  year={2022},
  publisher={APS}
}

@article{bechinger2016active,
  title={Active particles in complex and crowded environments},
  author={Bechinger, Clemens and Di Leonardo, Roberto and L{\"o}wen, Hartmut and Reichhardt, Charles and Volpe, Giorgio and Volpe, Giovanni},
  journal={Reviews of Modern Physics},
  volume={88},
  number={4},
  pages={045006},
  year={2016},
  publisher={APS}
}

@article{ray2023rectified,
  title={Rectified rotational dynamics of mobile inclusions in two-dimensional active nematics},
  author={Ray, Sattvic and Zhang, Jie and Dogic, Zvonimir},
  journal={Physical Review Letters},
  volume={130},
  number={23},
  pages={238301},
  year={2023},
  publisher={APS}
}

@article{houston2023colloids,
  title={Colloids in two-dimensional active nematics: conformal cogs and controllable spontaneous rotation},
  author={Houston, Alexander JH and Alexander, Gareth P},
  journal={New Journal of Physics},
  volume={25},
  number={12},
  pages={123006},
  year={2023},
  publisher={IOP Publishing}
}

@article{saw2018biological,
  title={Biological tissues as active nematic liquid crystals},
  author={Saw, Thuan Beng and Xi, Wang and Ladoux, Benoit and Lim, Chwee Teck},
  journal={Advanced Materials},
  volume={30},
  number={47},
  pages={1802579},
  year={2018},
  publisher={Wiley Online Library}
}

@article{woodhouse2013cytoplasmic,
  title={Cytoplasmic streaming in plant cells emerges naturally by microfilament self-organization},
  author={Woodhouse, Francis G and Goldstein, Raymond E},
  journal={Proceedings of the National Academy of Sciences},
  volume={110},
  number={35},
  pages={14132--14137},
  year={2013},
  publisher={National Academy of Sciences}
}

@article{aramini2025spontaneous,
  title={Spontaneous flow in active nematics: Effects induced by annular confinement},
  author={Aramini, A and Napoli, G and Turzi, S},
  journal={Physical Review E},
  volume={112},
  number={2},
  pages={025401},
  year={2025},
  publisher={APS}
}

@article{vaidya2024active,
  title={Active nematics in corrugated channels},
  author={Vaidya, Jaideep P and Shendruk, Tyler N and Thampi, Sumesh P},
  journal={Soft Matter},
  volume={20},
  number={41},
  pages={8230--8245},
  year={2024},
  publisher={Royal Society of Chemistry}
}

\end{document}